\documentclass[iop,revtex4]{emulateapj-rtx4}
\usepackage{graphicx}
\usepackage{rotating}
\usepackage{amsmath}
\usepackage[]{natbib}
\usepackage[usenames]{color} 
\usepackage{ulem}
\usepackage{lscape} 
\usepackage{rotating}
\newcommand{\mic}{$\mu$m}

\newcommand{\etal}{{et al.~}}
\newcommand\eg{{\it e.g.~}}

\shorttitle{2MASS light curves for SDSS$J015450+001501$: constraints for pulsation models}

\shortauthors{Szab\'{o} \etal}
\begin{document}

\title{High-precision 2MASS JHK$_s$ light curves and other data for RR Lyrae star SDSS$J015450+001501$:
            strong constraints for non-linear  pulsation models}

\author{R\'{o}bert Szab\'{o}\altaffilmark{1}}
\author{\v{Z}eljko Ivezi\'{c}\altaffilmark{1,2}}
\author{L\'{a}szl\'{o} L. Kiss\altaffilmark{1,3,4}}
\author{Zolt\'{a}n Koll\'{a}th\altaffilmark{1,5}}
\author{Lynne Jones\altaffilmark{2}}
\author{Branimir Sesar\altaffilmark{6}}
\author{Andrew C. Becker\altaffilmark{2}}
\author{James R.A. Davenport\altaffilmark{2}}
\author{Roc M. Cutri\altaffilmark{7}}

\altaffiltext{1}{Konkoly Observatory, MTA CSFK, Konkoly Thege Mikl\'{o}s \'{u}t 15-17, H-1121, 
                        Budapest, Hungary; {\tt rszabo@konkoly.hu} \label{Konkoly}}
\altaffiltext{2}{Astronomy  Department, University of  Washington, Box 351580, Seattle, WA 
                        98195-1580 \label{UW}}
\altaffiltext{3}{Sydney Institute for Astronomy, School of Physics, University of Sydney, NSW 2006, Australia}
\altaffiltext{4}{ELTE Gothard-Lend\"{u}let Research Group, Szent Imre herceg \'{u}t 112, H-9700 Szombathely, Hungary}
\altaffiltext{5}{University of West Hungary – Savaria Campus, Szombathely, Hungary}
\altaffiltext{6}{Division of Physics, Mathematics and Astronomy, Caltech, Pasadena, CA 91125 \label{Caltech}}
\altaffiltext{7}{Infrared Processing and Analysis Center, California Institute of Technology, Pasadena, 
                        CA 91125 \label{IPAC}}

\begin{abstract} 
We present and discuss an extensive data set for the non-Blazhko ab-type RR Lyrae star \objectname{SDSS$J015450+001501$},including optical SDSS $ugriz$ light curves and spectroscopic data, LINEAR and CSS 
unfiltered optical light curves, and infrared 2MASS JHK$_s$ and WISE W1 and W2 light curves. 
Most notably,  light curves obtained by 2MASS include close to 9000 photometric measures 
collected over 3.3 years and provide exceedingly precise view of near-IR variability. 
These data demonstrate that static atmosphere models are insufficient to explain
multi-band photometric light curve behavior and present strong constraints 
for non-linear pulsation models for RR Lyrae stars. It is a challenge to modelers to produce 
theoretical light curves that can explain data presented here, which we make publicly available. 
\end{abstract}


\section{Introduction}

RR Lyrae stars are old low-mass (around half the Sun's) pulsating horizontal branch stars of spectral 
class A and F. They are important both as stellar evolutionary probes and as tracers of Galactic
structure. For these reasons, accurate pulsational models for RR Lyrae stars are crucial in a
large number of astrophysical applications. \cite{Marconi2009} recently pointed out that a non-local 
time-dependent treatment of convection in non-linear (i.e., without small oscillation approximation)
pulsation models for RR Lyrae are needed to explain the morphological characteristics of the variation 
along a pulsation cycle of luminosity, radius, radial velocity, effective temperature and surface gravity:
``the comparison between theoretical and observed variations represents a powerful tool to constrain 
the intrinsic stellar parameters including the mass.'' 

Precise multi-band light curves provide strong constraints for pulsation models. \cite{Bono2000} performed 
simultaneous fitting of multi-band ($UBVK$) light curves of an RRc star (U Comae). They transformed 
the bolometric magnitudes supplied by a hydrocode to standard magnitudes using bolometric corrections and empirical 
color-temperature relations based on {\tt ATLAS9} model atmospheres \citep{Castelli1997}. 
\cite{Dorfi1999} carried out detailed frequency-dependent radiative transfer 
computations to obtain UBVI light curves of both RRab and RRc stars. Motivated by their results, 
in this work we present light curves for an RR Lyrae star pulsating in the fundamental mode (RRab) 
obtained over 15 years in ten photometric bandpasses that span more than a factor of ten in wavelength, and bracket the wavelength range where most of its luminosity is emitted. Most notably,  light curves obtained by 2MASS include close to 9000 photometric measures obtained over 3.3 years and, after data averaging, provide exceedingly precise view of near-IR variability. While quasi-continuous, extreme 
precision optical ligth curves have recently become available thanks to the $Kepler$ mission 
(\eg $\sim140000$ short cadence data points were analyzed for RRab stars FN Lyr and AW Dra in \citet{nemec2011} and half a million points per star were discussed in \citet{Nemec2013}), we believe that a data set of similar completeness, precision, and wavelength and temporal coverage does not exist for any other RR Lyrae star in the near-infrared bandpasses. Hence, data 
presented here will provide unprecedented observational constraints for non-linear pulsation 
models of RR Lyrae stars and will stimulate modelers to include a self-consistent and frequency-dependent 
treatment of radiation hydrodynamics in the outer parts of these pulsators.

In \S 2 we present and analyze the available data, which we make public, and discuss and
summarize our findings in \S 3.

\section{Data Analysis}

We first describe optical SDSS $ugriz$ light curves and spectroscopic data, 
LINEAR and CSS unfiltered light curve data, infrared 2MASS JHK$_s$ and WISE 
W1 and W2 light curve data, and then perform their joint analysis. The temporal 
coverage and sizes of these data sets are summarized in Table~\ref{table1}. 

As photometric data for RR\,Lyrae stars have been historically taken predominantly 
in the UBVRI system, it might be interesting to note that \cite{Ivezic2007} provide a 
set of non-linear transformation between BVRI and SDSS $griz$ magnitudes. For a linear 
transformation between the U and $u$ magnitudes the interested reader is referred to 
\cite{Jester2005}.

\begin{deluxetable}{lccrr}
\setlength{\tabcolsep}{0.02in} 
\tabletypesize{\scriptsize}
\tablecolumns{5}
\tablewidth{0pc}
\tablecaption{Photometric data analyzed in this work\label{table1}}
\tablehead{
\colhead{Survey} &
\colhead{MJD$_{\rm min}^a$} &
\colhead{MJD$_{\rm max}^b$} &
\colhead{N epochs$^c$} &
\colhead{N data$^d$} 
}
\startdata
2MASS    &    50660   &   51869    &    2969      &   8907   \\
SDSS       &    51075   &   54412    &        64      &    320   \\ 
LINEAR    &   52608   &   54817     &     225      &     225   \\
CSS         &    53627   &   56297    &      270     &      270  \\
WISE        &   55209    &  55575     &        43     &       86 \\
\enddata
\tablenotetext{a}{Earliest MJD in the survey.}
\tablenotetext{b}{Latest MJD in the survey.}
\tablenotetext{c}{The number of photometric epochs.}
\tablenotetext{d}{The number of photometric data points (the number of epochs times the number
of bandpasses).}
\end{deluxetable}

\subsection{SDSS data}

The Sloan Digital Sky Survey (SDSS; \citealt{SDSSDR7}) has repeatedly imaged
about 300 sq.~deg. large equatorial strip limited by $309\arcdeg < {\rm R.A.} < 60\arcdeg$ 
and $|{\rm Dec}|<1.26\arcdeg$, and known as Stripe 82. The properties of this data set
and its impact on variability studies are discussed in detail by \cite{Ivezic2007}, \cite{Sesar2007}
and \cite{Bramich2008}. An extensive study of light curves for RR Lyrae stars found in Stripe 82 was 
undertaken by \cite{Sesar2010}. 

The star discussed here was identified in SDSS Stripe 82 data with coordinates (J2000.0):
R.A. = 28.709054 and Dec= $+$0.250206 (decimal degrees). Following standard SDSS 
nomenclature, hereafter we refer to it as SDSS$J015450+001501$. All SDSS data for this
star are publicly\footnote{http://skyserver.sdss3.org/dr9/en/tools/explore/ \\
obj.asp?sid=788271533805037568} available, as well as published along with this paper 
electronically for convenience. Table~\ref{table6} shows the form and content of the data.
Images show a distinctively blue isolated 
point source, with the nearest brighter object more than 30 arcsec away. Its SDSS $r$ band 
magnitude varies from $\sim$14.6 to $\sim$15.5 and, according to \cite{Sesar2010}, this star 
is at a heliocentric distance of $D=7.59$ kpc.

\subsubsection{Imaging data}

\begin{figure}[!h]
\epsscale{1.2}
\vskip -0.70 in
\phantom{x} \hskip -0.0in
\plotone{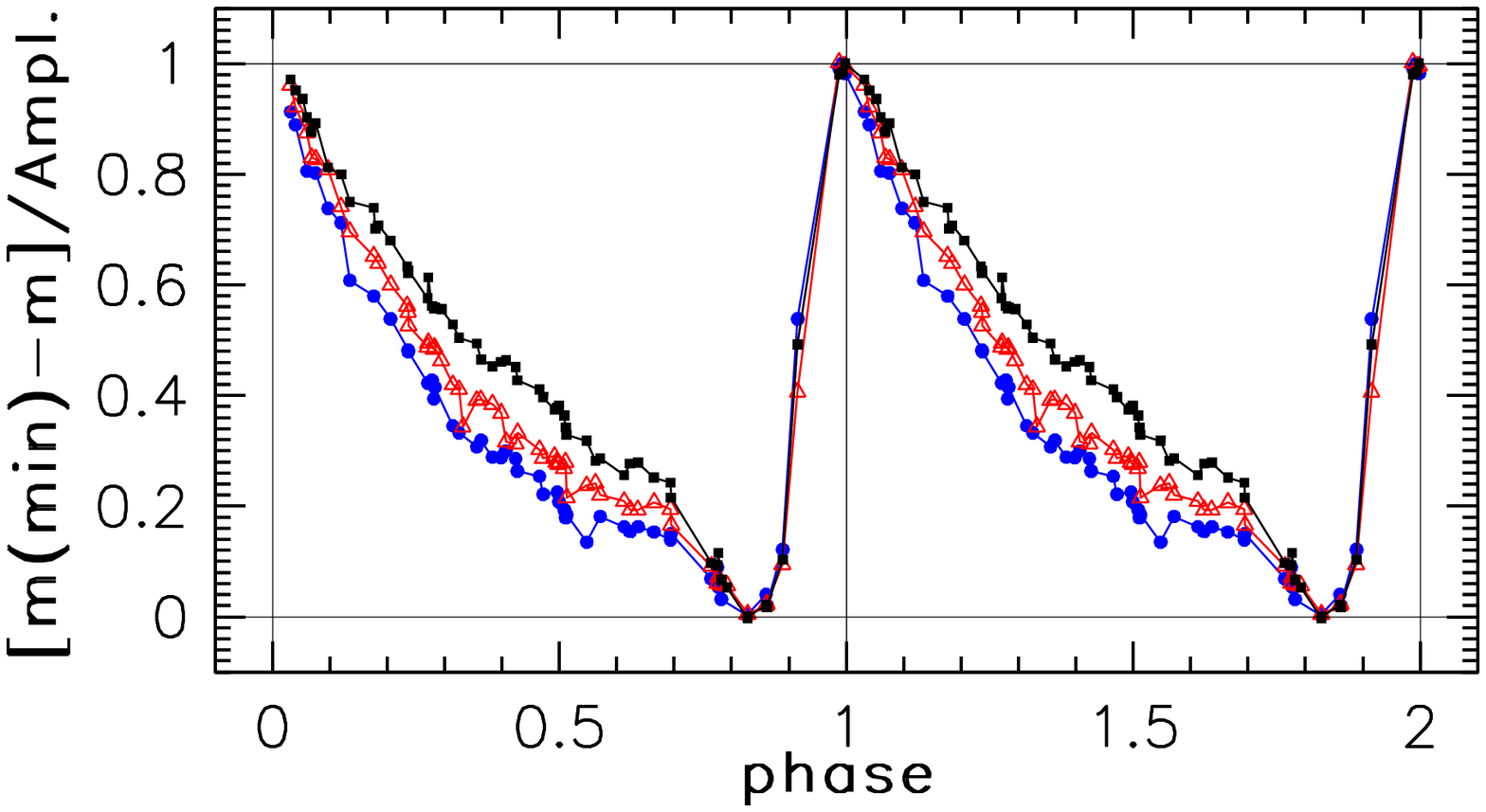}
\vskip -2.3 in
\caption{Phased and normalized SDSS $urz$ (bottom to top, respectively) light curves for 
SDSS$J015450+001501$. Data points are shown by symbols and connected by straight
lines. The scatter is due to random photometric errors (0.01-0.02 mag).}
\label{fig:SDSS}
\end{figure}

SDSS imaging data for this star were obtained 64 times and include nearly-simultaneous
$ugriz$ photometry precise to 0.01-0.02 mag. In order to construct phased light 
curves, we use the period and epoch of maximum determined using better sampled LINEAR data 
(see \S \ref{sec:LINEAR}). Phased and normalized SDSS $urz$ light curves are shown in 
Fig.~\ref{fig:SDSS}. Following \cite{Sesar2010}, we normalized light curves by their
amplitude (with minimum and maximum brightness determined using B splines)
and shifted so that minimum brightness is 0 and maximum brightness is 1. 
As evident from Fig.~\ref{fig:SDSS}, light curves are typical for ab RR Lyrae stars 
and greatly vary with wavelength.

\subsubsection{Spectroscopic data}

\begin{figure}[!h]
\epsscale{1.2}
\vskip -1.30 in
\phantom{x} \hskip -0.0in
\plotone{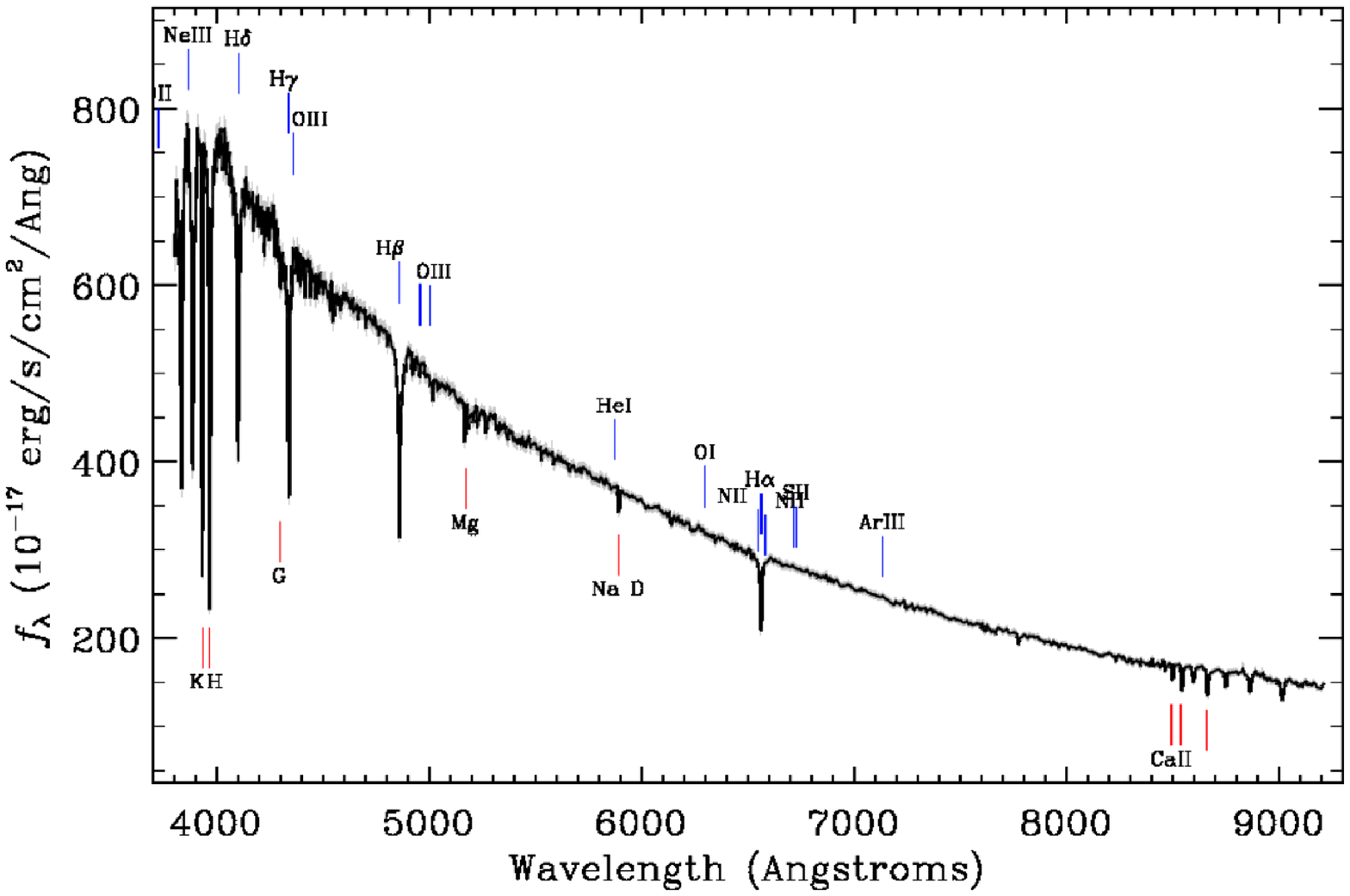}
\vskip -1.1 in
\caption{Default SDSS visualization of the SDSS spectrum for SDSS$J015450+001501$.
The spectrum is classified as an A0 star.}
\label{fig:spectrum}
\end{figure}

SDSS$J015450+001501$ was spectroscopically targeted by SDSS as a quasar candidate.  
Its SDSS spectrum (identifiers: plate=700, fiber=515, MJD=52199) is consistent with 
that for an A0 star (see Fig.~\ref{fig:spectrum}).  The spectroscopic parameters determined 
by the SEGUE Stellar Parameter Pipeline (\citealt{Lee2011segue}) include $[Fe/H]=-1.31\pm0.03$,
log(g)=2.97$\pm$0.20, T$_{\rm eff}$=6580$\pm$104 K, and radial velocity $-135.1$$\pm$1.8 km s$^{-1}$. 
We note however that the spectrum of this star is a coadd of 5 individual exposures, and the time 
between the start of the first and the end of the last exposure is 1.03 days. Thus, the parameters 
derived from this spectrum might contain much higher systematic errors because the individual 
exposures span a wide range of pulsational phases.

\subsection{LINEAR and CSS light curve data}
\label{sec:LINEAR}

\begin{deluxetable}{ccc}
\tablecolumns{3}
\tablewidth{7cm}
\tablecaption{LINEAR light curve for RR\,Lyrae star SDSS$J015450+001501$ (225 epochs)\label{table2}}
\tablehead{
\colhead{MJD} &
\colhead{mag} &
\colhead{mag$_{\rm err}$} 
}
\startdata
52608.145642 &   15.559  &   0.257 \\
52608.175414 &   15.424  &   0.024 \\
... & ... & ... 
\enddata
\tablecomments{For details see \cite{Sesar2011, Sesar2013}. Table~\ref{table2} is published in its entirety in the 
electronic edition of the Journal. A portion is shown here for guidance regarding its form and content.}
\end{deluxetable}

The asteroid survey LINEAR was photometrically recalibrated by \cite{Sesar2011} 
using SDSS stars acting as a dense grid of standard stars. In the overlapping
$\sim$10,000 deg$^2$ of sky between LINEAR and SDSS, photometric errors for 
unfiltered (white light) photometry range from $\sim$0.03 mag for sources not limited by photon
statistics to $\sim$0.20 mag at $r=18$ (here $r$ is the SDSS $r$ band magnitude). 
LINEAR data provide time domain information for the brightest 4 magnitudes of 
SDSS survey, with 250 unfiltered photometric observations per object on average.
The public access to the recalibrated LINEAR data is provided through the SkyDOT
Web site (https://astroweb.lanl.gov/lineardb/) and is also available as auxiliary material 
with this paper (see Table~\ref{table2}). RR Lyrae stars from this data set 
have been analyzed by \cite{Sesar2013}. 

The Catalina Sky Survey (CSS) uses three telescopes to search for near-Earth objects.
Each of the survey telescopes is run as separate sub-surveys, including the Catalina 
Schmidt Survey and the Mount Lemmon Survey in Tucson, Arizona, and the Siding
Spring Survey in Siding Spring, Australia. CSS is similar to LINEAR in that it uses
unfiltered observations and delivers similar photometric precision (but is deeper by 
several magnitudes). RR Lyrae stars from this data set have been analyzed 
by \cite{Drake2013}. 

\begin{figure}[!h]
\epsscale{1.2}
\vskip -0.70 in
\phantom{x} \hskip -0.0in
\plotone{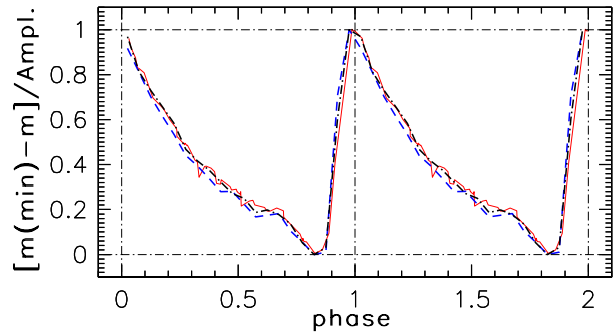}
\vskip -2.3 in
\caption{LINEAR (dashed), CSS (dot-dashed) and SDSS $r$-band (solid) light curves for 
SDSS$J015450+001501$. LINEAR and CSS data (225 and 270 data points, respectively)
are median-ed in 0.05 wide phase bins, with resulting random errors at most 0.01 mag. 
SDSS light curve corresponds to 64 data points (with random errors of 0.01-0.02 mag)
connected by straight lines (same curve as in Figure~\ref{fig:SDSS}). Note that all three light 
curves barely differ from each other.}
\label{fig:LINEARlc}
\end{figure}

SDSS$J015450+001501$ was observed by LINEAR and CSS about 500 times over 10 years. 
Both light curves are shown in Fig.~\ref{fig:LINEARlc} and are barely distinguishable. 
Using only LINEAR data, \cite{Sesar2013} have determined a best-fit period,
$P=0.636985$ day, and the epoch of maximum, MJD$_{\rm max}$ = 53675.299080.
Based on \cite{Sesar2011} photometric transformations between LINEAR and SDSS
$r$-band data, a great degree of similarity is expected between unfiltered LINEAR/CSS
light curves and the SDSS $r$-band light curve. This expectation is verified by data, 
as illustrated in Fig.~\ref{fig:LINEARlc}. CSS observations of SDSS$J015450+001501$ 
are collected in Table~\ref{table3}. 

\begin{deluxetable}{ccccccc}
\tablecolumns{3}
\tablewidth{7cm}
\tablecaption{CSS light curve for RR\,Lyrae star SDSS$J015450+001501$ (270 epochs)\label{table3}}
\tablehead{
\colhead{MJD} &
\colhead{mag} &
\colhead{mag$_{\rm err}$}
}
\startdata
53710.18424 & 15.44  & 0.06 & \\
53710.19176 & 15.46 &  0.06 & \\
... & ... & ... 
\enddata
\tablecomments{Table~\ref{table3} is published in its entirety in the electronic edition of the Journal. A portion is shown here for guidance regarding its form and content.}
\end{deluxetable}

\subsection{2MASS light curve data} 

The most extensive data set presented here comes from the 2MASS survey (\citealt {Skrutskie2006}). 
SDSS$J015450+001501$ falls in one of the 35 2MASS calibration fields that were imaged repeatedly 
in the J, H and K$_s$ bandpasses during each night of the 3.5 year survey. Full details are given in 
the online Explanatory Supplement\footnote{http://www.ipac.caltech.edu/ \\ 2mass/releases/allsky/doc/sec3\_2d.html} 
and \cite{Cutri2003}. Analysis of various variable stars contained in this data set was presented
by \cite{Plavchan2008}, \cite{Becker2008} and \cite{Davenport2012}, where more technical details 
can be found. 

\begin{figure}[!h]
\epsscale{1.2}
\vskip -0.70 in
\phantom{x} \hskip -0.0in
\plotone{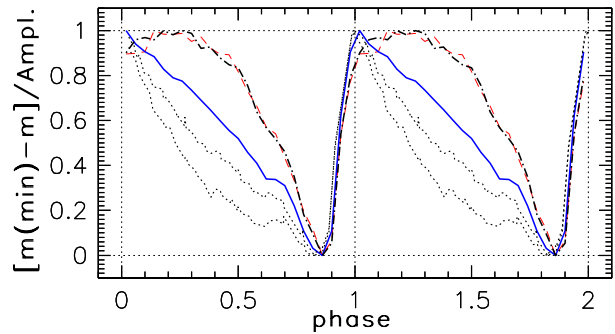}
\vskip -2.3 in
\caption{Phased and normalized SDSS $gi$ and 2MASS $JHK$ (bottom to top, respectively) light curves
for SDSS$J015450+001501$. Phased 2MASS light curve data ($\sim$3000 data points per band) 
are median-ed in 0.04 wide phase bins, with resulting random errors well below 0.01 mag. SDSS light curves
correspond to 64 data points each (with random errors of 0.01-0.02 mag), connected by straight lines.
Note that the shapes of $H$ and $K$ light curves are barely distinguishable.}
\label{fig:2MASS}
\end{figure}

SDSS$J015450+001501$ was observed about 3,000 times in each of the three bandpasses (see
Tables~\ref{table1} \& \ref{table8}). 
Phased and normalized 2MASS light curves are shown and compared to SDSS $gi$ light curves in
Fig.~\ref{fig:2MASS}. Because the 2MASS data set is so large and we see no sign of Blazhko-modulation 
in the light curves, data are median-ed in 0.04 wide phase bins, with resulting random errors well 
below 0.01 mag. The variation of the light curve shape with bandpass wavelength seen in SDSS data 
(see Fig.~\ref{fig:SDSS}), continues into near-IR, but only to the $H$ band -- the light curves in 
the $H$ and $K$ bands are barely distinguishable despite the high precision of 2MASS data. We proceed 
with a more detailed analysis of the $JHK$ variability which reveals interesting light curve features. 

\subsubsection{Phase-resolved near-IR color-magnitude hysteresis loops}

\begin{figure}[!h]
\epsscale{1.24}
\phantom{x}\vskip -0.5in
\phantom{x}\hskip -0.2in
\plotone{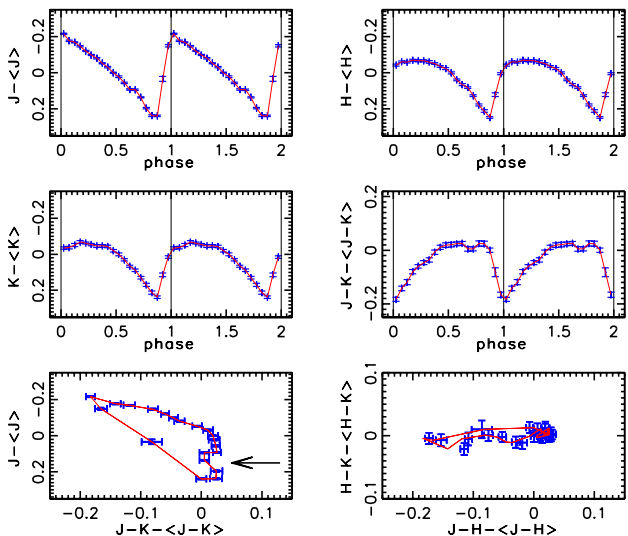}
\vskip -1.7 in
\caption{Top four panels display phased and normalized 2MASS $JHK$ and color $J-K$ 
light curves for SDSS$J015450+001501$. Phased 2MASS light curve data ($\sim$3000 data 
points per band) are median-ed in 0.05 wide phase bins, with resulting random errors well 
below 0.01 mag. The median value of these averaged data points is subtracted from each curve, 
and points are connected by 
straight lines. Random errors are derived from scatter in each phase bin. Bottom two
panels show color-magnitude (left) and color-color (right) hysteresis curves (the motion
is clockwise). Note that between phases 0.65 and 0.70 the $J-K$ color becomes bluer 
despite the decreasing brightness (see the middle right and bottom left panels,
and arrow in the latter panel). The $H-K$ color does not show any variation (rms of 
$\sim0.01$ mag, consistent with measurement errors).}
\label{fig:JHKlightcurves}
\end{figure}

The phased $JHK$ light curves, the $J-K$ color variation with phase, and phased-resolved
color-magnitude and color-color diagrams constructed with median-ed 2MASS data are 
shown in Fig.~\ref{fig:JHKlightcurves}. A notable feature is that the time of maximum light 
in $H$ and $K$ bandpasses does not coincide with the time of maximum light at 
shorter wavelengths, but lags in phase by about 0.25. Unlike the $J-K$ color which varies 
with an amplitude of $\sim$0.2 mag, the $H-K$ color does not appear to vary at all: its 
root-mean-square scatter is only 0.01 mag, and consistent with photometric noise 
(compare to 0.075 mag for the $J-K$ color). 

Perhaps the most interesting feature revealed by the high-precision 2MASS data
is seen in the $J$ vs. $J-K$ color-magnitude diagram (bottom left panel of Fig.~\ref{fig:JHKlightcurves}). 
In addition to the clearly seen ``hysteresis'' (for illustration of a similar behavior in the $ugr$ bandpasses, 
see Figure~9 in \citealt{Sesar2010}), the $J-K$ color becomes bluer between phases 0.65 and 0.70 
despite the decreasing brightness (see the middle right and bottom left panels,
and arrow in the latter panel). Given the short duration ($\sim$45 minutes) and a small 
amplitude (0.03 mag) of this feature, it is unlikely that it was well observed for other stars.
We return to an interpretation of this behavior in \S\ref{sec:sim}.

\subsection{WISE light curve data}

\begin{figure}[!h]
\epsscale{1.2}
\vskip -0.70 in
\phantom{x} \hskip 0.1in
\plotone{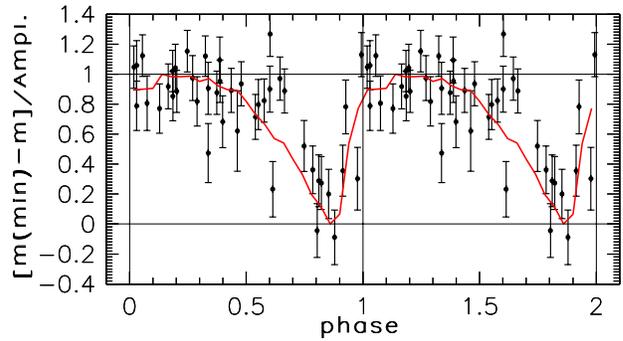}
\vskip -2.3 in
\caption{WISE $W1$-band phased and normalized light curve for SDSS$J015450+001501$
is shown as symbols with error bars (as computed by the WISE photometric processing pipeline). The line
shows phased and normalized 2MASS $K$-band light curve (same curve as in Figure~\ref{fig:2MASS}).}
\label{fig:WISE}
\end{figure}

\begin{deluxetable}{ccccc}
\tablecolumns{5}
\tablewidth{7cm}
\tablecaption{WISE light curves for RR\,Lyrae star SDSS$J015450+001501$ (43 epochs)\label{table4}}
\tablehead{
\colhead{MJD} &
\colhead{W1} & \colhead{W1$_{\rm err}$} &
\colhead{W2} & \colhead{W2$_{\rm err}$} 
}
\startdata
55209.272845 & 13.853 & 0.047 & 14.016 & 0.126 \\
55209.405150 & 13.828 & 0.052 & 13.770 & 0.119 \\
... & ... & ... & ... & ... \\
\enddata
\tablecomments{Table~\ref{table4} is published in its entirety in the electronic edition of the Journal. A portion is shown here for guidance regarding its form and content.}
\end{deluxetable}

The Wide-field Infrared Survey Explorer (WISE, \citealt{Wright2010})  mapped the sky at 3.4, 
4.6, 12, and 22 \mic. WISE imaged each point on the sky multiple times to achieve its sensitivity
goals and to reject transient events such as cosmic rays.  SDSS$J015450+001501$ 
was measured 43 independent times, during two epochs separated by 
approximately six months. The single-exposure photometric signal-to-noise ratio is high enough for light curve
analysis in the W1 (3.4 \mic) and W2 (4.6 \mic) bands: the median photometric
errors are 0.05 mag for W1 and 0.15 mag for W2. These two light curves
(43 data points per band) are compiled in Table~\ref{table4}. 

Although WISE data are noisy, it is possible to make a rough comparison with 2MASS data.
As shown in Fig.~\ref{fig:WISE}, the phased and normalized light curve in the W1 band is fully consistent 
with the corresponding 2MASS light curve in the $K$ band (including the amplitude). A more detailed 
comparison of data across all ten bandpasses is described next.

\subsection{Simultaneous analysis of optical and infrared data} 
\label{sec:sim}

\begin{figure}[!h]
\epsscale{1.2}
\vskip -0.70 in
\phantom{x} \hskip -0.0in
\plotone{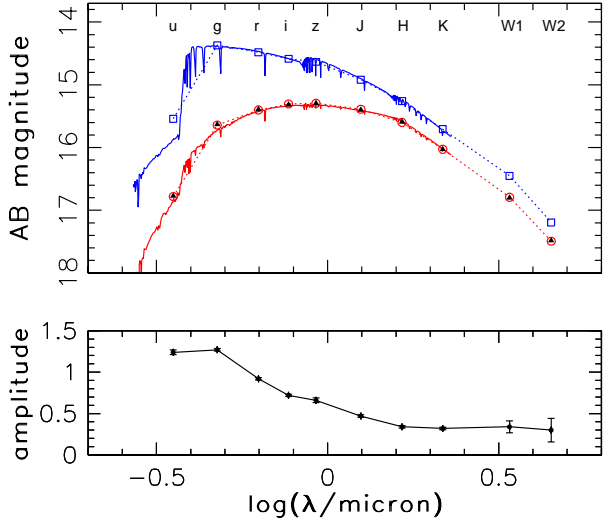}
\vskip -1.3 in
\caption{Top panel: ten-band spectral energy distributions (SED) at maximum (top) and minimum (bottom) 
light for SDSS$J015450+001501$. Data are shown by symbols and connected by dotted straight lines. Open
squares correspond to maximum light and open circles to minimum light (phase=0.86). Solid triangles,
inside open circles, correspond to phase=0.50, {\it shifted fainter} by 0.25 mag. The SED {\it shape} at
phase=0.50 is nearly indistinguishable from the SED at minimum light. The solid lines show Kurucz models 
for $[Fe/H] = -1.5$, log(g)=3.0, and $T_{\rm eff}=7500$ K (top) and $T_{\rm eff}=6000$ K (bottom). Bottom 
panel: light curve amplitude, in magnitudes, as a function of wavelength.}
\label{fig:SEDampl}
\end{figure}

As shown above, the shape of light curves varies greatly with wavelength. Another way 
to look at the same data set is to construct phase-resolved spectral energy distributions (SED).
SEDs at minimum and maximum phases, and at phase=0.5, are compiled in Table~\ref{table5} and 
shown in the top panel in 
Fig.~\ref{fig:SEDampl}. When the SED at phase=0.5 is scaled to a fainter flux level by 0.25 mag,
it is indistinguishable from the SED at minimum phase (0.86). Assuming identical effective
temperatures, this scaling factor implies that the effective radius of the star decreases 
by 12\% between these two phases. 

The variation of light curve amplitude with wavelength is shown in the bottom panel in 
Fig.~\ref{fig:SEDampl}. Between the SDSS $g$ band (0.476 \mic) and the 2MASS $H$ band
(1.65 \mic), the amplitude steadily decreases from $\sim$1.3 mag to $\sim$0.3 mag. 
At longer wavelengths there is very little change of light curve amplitude, if any. 
The $u$ band and the $g$ band amplitudes are similar, which implies that the $u-g$
color stays approximately constant over the pulsational cycle. This is easily understood
as the consequence of the fact that the $u-g$ color primarily depends on metallicity 
(see \citealt{Marconi2006}).  

The overall SED shapes are well described by static model atmospheres computed by \cite{Kurucz1979}. 
The solid lines in the top panel in Fig.~\ref{fig:SEDampl} show Kurucz models for fixed $[Fe/H] = -1.5$ 
and log(g)=3.0, and best-fit $T_{\rm eff}=7500$ K (maximum light) and $T_{\rm eff}=6000$ K (minimum 
light). The effective temperature step in the model library was 250 K. A slight discrepancy between
the measured $u$ band magnitude at maximum light and the best-fit model can be reconciled as due
to the impact of the steep SED in that region. 

\begin{deluxetable*}{ccccccccc}
\tablecolumns{9}
\tablewidth{0pc}
\tablecaption{Spectral energy distribution data for RR\,Lyrae star SDSS$J015450+001501$\label{table5}}
\tablehead{
\colhead{band} & \colhead{$\lambda_{\rm eff}$ (${\rm \mu m}$)} & \colhead{mag$_{\rm min}$} & \colhead{err} &
\colhead{mag$_{\rm max}$} & \colhead{err} &  \colhead{amplitude} & \colhead{mag$_{\phi = 0.5}$} & \colhead{extinction corr.}}
\startdata
   u   &  0.3540  &    16.96 &  0.02 & 15.72 &  0.02  &   1.24   &      16.70  &      0.179 \\
   g   &  0.4760  &    15.78 &  0.01 & 14.51 &  0.01  &   1.27   &      15.52  &      0.139 \\
   r   &  0.6280  &    15.50 &  0.01 & 14.58 &  0.01  &   0.92   &      15.24  &      0.099 \\
   i   &  0.7690  &    15.38 &  0.01 & 14.66 &  0.01  &   0.72   &      15.14  &      0.075 \\
   z   &  0.9250  &    15.35 &  0.02 & 14.69 &  0.02  &   0.66   &      15.10  &      0.056 \\
   J   &   1.25   &    15.42 &  0.01 & 14.95 &  0.01  &   0.47   &      15.18  &      0.031 \\
   H   &   1.65   &    15.62 &  0.01 & 15.28 &  0.01  &   0.34   &      15.36  &      0.020 \\
   K   &   2.17   &    16.04 &  0.01 & 15.72 &  0.01  &   0.32   &      15.79  &      0.013 \\
  W1   &   3.4    &    16.80 &  0.05 & 16.46 &  0.05  &   0.34   &      16.56  &      0.008 \\ 
  W2   &   4.5    &    17.50 &  0.10 & 17.20 &  0.10  &   0.30   &      17.24  &      0.007 
\enddata
\tablecomments{The table lists 2MASS and WISE magnitudes on AB scale. The following Vega-to-AB magnitude offsets 
(m$_{\rm \,AB}$=m$_{\rm \,Vega}$+offset) were used: 0.89, 1.37, 1.84 in $JHK$, and 2.6, 3.3 in $W1$ and $W2$. 
Magnitude values are not corrected for ISM dust extinction. 
Plausible extinction corrections \citep{Schlegel1998} are listed in the last column and are derived using the SFD value for the $r$-band 
extinction quoted by SDSS for this star: A$_r$ = 0.099, and coefficients listed in the first row of Table 1
from \cite{Berry2012} for the $ugrizJHK$ bands. For the WISE bands, the coefficients taken from 
\cite{Yuan2013} (A$_{\rm W1}$/A$_{\rm r}$ = 0.084, A$_{\rm W2}$/A$_{\rm r}$ = 0.074).}
\end{deluxetable*}

\begin{figure}[!h]
\epsscale{1.2}
\vskip -0.70 in
\phantom{x} \hskip -0.0in
\plotone{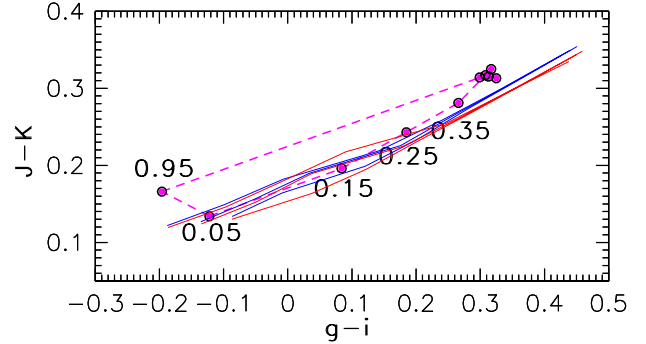}
\vskip -2.3 in
\caption{The symbols connected by dashed lines show 2MASS $J-K$ color vs. SDSS $g-i$
color for SDSS$J015450+001501$. Data were binned in 0.1 wide equidistant phase bins,
and the bin centers are marked for a few data points (the motion during pulsational cycle is
counter-clockwise in this diagram). Although both colors are good estimators of the effective
temperature, a hysteresis is easily seen (e.g., at $g-i=0$, about 0.05 mag difference in
$J-K$ color). Note that both colors are essentially constant for phase bins from 0.45 to 085.
The six thin solid lines are predictions based on Kurucz static model atmospheres with
effective temperatures in the range 6000--7500 K. At the blue end, they bifurcate into
the groups of three tracks -- they correspond to models with $[Fe/H] = -1.5$ (the top
three curves) and $[Fe/H] = -1.0$. For each metallicity, the three curves correspond
to log(g)=2.5, 3.0 and 3.5. Note that at the blue end, the models bifurcate according
to log(g), rather than $[Fe/H]$ (at $J-K\sim 0.13$, the $g-i$ color becomes bluer by
$\sim$0.1 mag as log(g) increases from 2.5 to 3.5.
}
\label{fig:CC}
\end{figure}

Despite this apparent success of static atmosphere models, it is easy to demonstrate that
they cannot provide a complete description of the data. Fig.~\ref{fig:CC} shows a phase-resolved
$J-K$ vs. $g-i$ color-color diagram. Both of these colors are by and large driven by effective
temperature and Kurucz models show that the impact of log(g) is only minor (at most a few
hundredths of a magnitude in the $J-K$ color at a given $g-i$ color, for a detailed discussion
see \citealt{Ivezic08}). Therefore, if static atmosphere models are sufficient to describe the
data, then the data should follow a single line in this diagram, with the position along that line
controlled by the effective temperature. In contrast, data in this color-color diagram show a
similar hysteresis effect as the $J$ vs. $J-K$ color-magnitude diagram (bottom left panel in
Fig.~\ref{fig:JHKlightcurves}). This color-color hysteresis, however,  cannot be described as
due to different stellar radii at a given effective temperature, as in the case of color-magnitude
diagram.

To provide a precise quantitative estimate of the discrepancy between observed colors
and colors predicted by Kurucz models, we compare a set of models for three values of
log(g) and two values of $[Fe/H]$ that bracket the values expected for this star (see
Fig.~\ref{fig:CC}). Models provide a fairly good agreement, to within a few hundredths of a
magnitude) for the descending branch of the light curve (phases between the light
curve maximum and minimum, the lower branch in Fig.~\ref{fig:CC}). However, all models
fail to explain the observed colors during the ascending branch, as the light curve goes
from the minimum light to the maximum light (note that ascending portion of the light curve
is about five times shorter than the descending portion). At a given $J-K$ colors, the
discrepancy can be as large as $\sim$0.2 mag!

Note that this model failing cannot be simply attributed to inadequacy of Kurucz models:
whatever static atmosphere model library is invoked, it should produce much smaller hysteresis
in this diagram, if any. For example, at a fixed $J-K=0.2$, the difference in the $g-i$ color at the
descending and ascending branches is over 0.2 mag. Similarly, at a fixed $g-i=0$ color, the
$J-K$ color differs by $\sim$0.05 mag. Such large color differences are unlikely to be explained
by the impact of log(g) parameter on static atmosphere models. It remains to be seen whether
dynamic models can explain these data.

\section{Discussion and Conclusions} 

Essentially by coincidence, the non-modulated (non-Blazhko) RR Lyrae star SDSS$J015450+001501$ has been 
extensively observed by several major sky surveys: SDSS, 2MASS, CSS, LINEAR and WISE. We have compiled 
data from these surveys and made the compilation publicly available in a user-friendly form. The light 
curves for this star are available in ten photometric bandpasses that span more than a factor of ten in 
wavelength and bracket the wavelength range where most of its luminosity is emitted. Light curves obtained by 
2MASS stand out: they include close to 9000 photometric measures and provide an unprecedentedly 
precise view of near-IR variability. A data set of similar completeness, precision, and wavelength and 
temporal coverage likely does not exist for any other RR Lyrae star. 

The data for SDSS$J015450+001501$ presented here provide strong observational constraints for 
non-linear pulsation models of RR Lyrae stars. \cite{Sesar2010} already pointed out (see their Figure 8)
that there are discrepancies between the $ugriz$ RR Lyrae light curves obtained by SDSS and theoretical 
light curve predictions from \cite{Marconi2006}. Here, we further demonstrated that static atmosphere 
models are insufficient to explain multi-band photometric light curve behavior. Indeed, \cite{Barcza2010} 
demonstrated that the quasi-static approximation is not valid for all phases during an RR\,Lyrae 
pulsational cycle. Pulsation models computed by \cite{Fokin1997} with a large number of mass shells in the 
stellar atmosphere show that the s3' and/or the merging s3+s4 shock waves might be at work at the pulsational 
phase range of 0.65-0.70, where we found the 'kink' in the $J-K$ color progression (Fig~\ref{fig:JHKlightcurves}). 
We caution however that a large model survey should be conducted in a broad range of RR\,Lyrae physical parameters 
in order to verify the existence and the influence of these shock waves, in particular in SDSS$J015450+001501$.

Another complication is that in different passbands we observe the integrated light 
of different atmospheric layers. Therefore a frequency-dependent, dynamical atmosphere model coupled to a 
state-of-the-art hydrocode is needed to consistently handle the violent, shock-permeated outer parts of 
such large-amplitude pulsators. Frequency-dependent treatment of radiation hydrodynamics developed by \cite{Dorfi1999} 
is very promising in this respect. They found good agreement between the synthetic optical RR\,Lyrae light curves and 
observations. An extension of such calculations to the near infrared passbands would be most beneficial for comparison 
with the data presented here. 


\section{Acknowledgments}
\label{s:ack}
\v{Z.} Ivezi\'{c} thanks the Hungarian Academy of Sciences for its support through the 
Distinguished Guest Professor grant No. E-1109/6/2012.  ACB and JRAD acknowledge support 
from NASA ADP grant NNX09AC77G. This project has been supported by 
the `Lend\"ulet-2009 Young Researchers' Program of the Hungarian Academy of Sciences, the 
HUMAN MB08C 81013 grant of the MAG Zrt, the Hungarian OTKA grant K83790 and the 
KTIA URKUT\_10-1-2011-0019 grant. R. Szab\'{o} was supported by the J\'anos Bolyai Research 
Scholarship of the Hungarian Academy of Sciences. 

This publication makes use of data products 
from the Two Micron All Sky Survey, which is a joint project of the University of Massachusetts 
and the Infrared Processing and Analysis Center/California Institute of Technology, funded by 
the National Aeronautics and Space Administration and the National Science Foundation. This publication 
also makes use of data products from the Wide-field Infrared Survey Explorer and NEOWISE which are joint 
projects of the University of California, Los Angeles, and the Jet Propulsion Laboratory/California 
Institute of Technology, funded by the National Aeronautics and Space Administration. 

Funding for the 
SDSS and SDSS-II has been provided by the Alfred P. Sloan Foundation, the Participating Institutions, 
the National Science Foundation, the U.S. Department of Energy, the National Aeronautics and Space 
Administration, the Japanese Monbukagakusho, the Max Planck Society, and the Higher Education Funding 
Council for England. The SDSS Web Site is http://www.sdss.org/.
The SDSS is managed by the Astrophysical Research Consortium for the Participating Institutions. 
The Participating Institutions are the American Museum of Natural History, Astrophysical Institute 
Potsdam, University of Basel, University of Cambridge, Case Western Reserve University, University 
of Chicago, Drexel University, Fermilab, the Institute for Advanced Study, the Japan Participation 
Group, Johns Hopkins University, the Joint Institute for Nuclear Astrophysics, the Kavli Institute 
for Particle Astrophysics and Cosmology, the Korean Scientist Group, the Chinese Academy of Sciences 
(LAMOST), Los Alamos National Laboratory, the Max-Planck-Institute for Astronomy (MPIA), the 
Max-Planck-Institute for Astrophysics (MPA), New Mexico State University, Ohio State University, 
University of Pittsburgh, University of Portsmouth, Princeton University, the United States Naval 
Observatory, and the University of Washington.

\bibliography{ref}
\bibliographystyle{aj}


\clearpage

\begin{deluxetable}{ccrcccccccc}
\setlength{\tabcolsep}{0.02in} 
\tabletypesize{\scriptsize}
\tablecolumns{11}
\tablewidth{0pc}
\tablecaption{SDSS $ugriz$ light curves for RR Lyrae star SDSS$J015450+001501$ (64 epochs)}
\tablehead{
\colhead{R.A.$^a$} & \colhead{Dec$^a$} &
\colhead{MJD$_{u}$} & \colhead{$u$} & \colhead{$u_{\rm err}$} &
\colhead{MJD$_{g}$} & \colhead{$g$} & \colhead{$g_{\rm err}$} &
\colhead{MJD$_{r}$} & \colhead{$r$} & \colhead{$r_{\rm err}$} \\
\colhead{(deg)} & \colhead{(deg)} & 
\colhead{(day)} & \colhead{(mag)} & \colhead{(mag)} & 
\colhead{(day)} & \colhead{(mag)} & \colhead{(mag)} & 
\colhead{(day)} & \colhead{(mag)} & \colhead{(mag)}}
\startdata
28.709058 & 0.250204 & 51075.379381 & 16.603 & 0.008 & 51075.381047 & 15.412 & 0.005 & 51075.377714 & 15.161 & 0.006 \\
28.709058 & 0.250204 & 51818.349055 & 16.921 & 0.009 & 51818.350721 & 15.715 & 0.006 & 51818.347388 & 15.447 & 0.006  \\
... & ... & ... & ... & ... & ... & ... & ... & ... & ... & ... 
\enddata
\tablenotetext{a}{Equatorial J2000.0 right ascension and declination.}
\tablecomments{Magnitudes are not corrected for ISM dust extinction and set to -99.999 if unreliable. See \cite{Sesar2010} for more details. 
Table~\ref{table6} is published in its entirety  (including the $iz$ photometry) in the electronic edition of the Journal. A portion is shown here for guidance regarding its form and content.}
\label{table6}
\end{deluxetable}

\begin{deluxetable}{ccccccccccccc}
\setlength{\tabcolsep}{0.02in} 
\tabletypesize{\scriptsize}
\tablecolumns{13}
\tablewidth{0pc}
\tablecaption{2MASS $JHK$ light curves for RR Lyrae star SDSS$J015450+001501$ (2969 epochs)}
\tablehead{
\colhead{SourceID} & 
\colhead{R.A.$^a$} & \colhead{Dec$^a$} & \colhead{BJD} &
\colhead{$J$} & \colhead{$J_{\rm err}$} & \colhead{$J_{\rm q}$$^b$} &
\colhead{$H$} & \colhead{$H_{\rm err}$} & \colhead{$H_{\rm q}$$^b$} &
\colhead{$K$} & \colhead{$K_{\rm err}$} & \colhead{$K_{\rm q}$$^b$} \\
\colhead{} & \colhead{(deg)} & \colhead{(deg)} & \colhead{} &
\colhead{(mag)} & \colhead{(mag)} & \colhead{} & 
\colhead{(mag)} & \colhead{(mag)} & \colhead{} &
\colhead{(mag)} & \colhead{(mag)} & \colhead{} }
\startdata
10038165 & 28.709000 & 0.250413 & 51003.4463211 & 14.257 & 0.030 & A & 14.007 & 0.041 & A & 13.947 & 0.075 & A \\
10038255 & 28.708985 & 0.250394 & 51003.4466211 & 14.219 & 0.031 & A & 13.949 & 0.040 & A & 13.941 & 0.077 & A \\
... & ... & ... &  ... & ... & ... &... & ... & ... &... & ... & ... & ... \\
\enddata
\tablenotetext{a}{Equatorial J2000.0 right ascension and declination.}
\tablenotetext{b}{Reliability flag. Single character flag that is related to the probability (P) that the extraction is a valid detection of a near infrared source on the sky at the time of the observation. "A" means P$>90$\%.}
\tablecomments{Table~\ref{table8} is published in its entirety in the electronic edition of the Journal. A portion is shown here for guidance regarding its form and content.}
\label{table8}
\end{deluxetable}

\end{document}